\documentclass[conference]{IEEEtran}
\IEEEoverridecommandlockouts
\usepackage{cite}
\usepackage{amsmath,amssymb,amsfonts}
\usepackage{algorithmic}
\usepackage{graphicx}
\usepackage{textcomp}
\usepackage{xcolor}
\def\BibTeX{{\rm B\kern-.05em{\sc i\kern-.025em b}\kern-.08em
    T\kern-.1667em\lower.7ex\hbox{E}\kern-.125emX}}
\begin{document}

\title{FDTD simulation algorithm of magnetized plasma media}

\author{\IEEEauthorblockN{Aleksandr Friman}
\IEEEauthorblockA{\textit{Solid State Physics department} \\
\textit{LPI RAS}\\
Moscow, Russia \\
friman\_a@sci.lebedev.ru}
}

\maketitle

\begin{abstract}
A technique of magnetized plasma simulation have been implemented into a well known FDTD solver MEEP as a child class of "susceptibility" class. Magnetized plasma posses gyrotrophic properties, 
polarization vector is being rotated while the wave propagates through the material. Solid state plasma can be found in semiconductor materials which rises a question of using gyrotrophic 
properties of plasma in modern photonic development. 
To test the software, a circular polarized wave propagation through 9mm layer of magnetized plasma has been simulated and compared to the analytical solution.
\end{abstract}

\begin{IEEEkeywords}
electrodynamics, finite difference methods, gyrotropism, microwave photonics, magnetized plasma
\end{IEEEkeywords}

\section{Introduction}
Plasma media are widespread in solid state physics. These media can be magnetized and behave as magnetoactive materials in electromagnetic systems. 
One of important properties of magnetized plasmas is Parity Time symmetry (PT-symmetry) breaking. That leads to many interesting phenomena such as a new class of Talbot effects \cite{Talbot},
unidirectional invisibility \cite{inv}\cite{inv2} 
and absorption \cite{abs} control.
For semiconductor industry 
especially important kind of plasmas for semiconductor industry is solid state plasmas. 
Using magnetized solid state plasma medias would allow to exploit PT symmetry breaking effect in integrated circuits. 
Apparently we have found out that there was no publicly available FDTD software capable to simulate electromagnetic effects in magnetized plasmas. 
To overcome this problem we initially modified MEEP to 
simulate magnetized plasmas with cyclotron resonance caused by Z-axle directed static magnetic field \cite{our_paper1}. In this work we are presenting a next version of the 
algorithm allowing to set any direction of the static magnetic field.

\section{FDTD basics}
\begin{figure}[h]
\center{\includegraphics[width=1.0\linewidth]{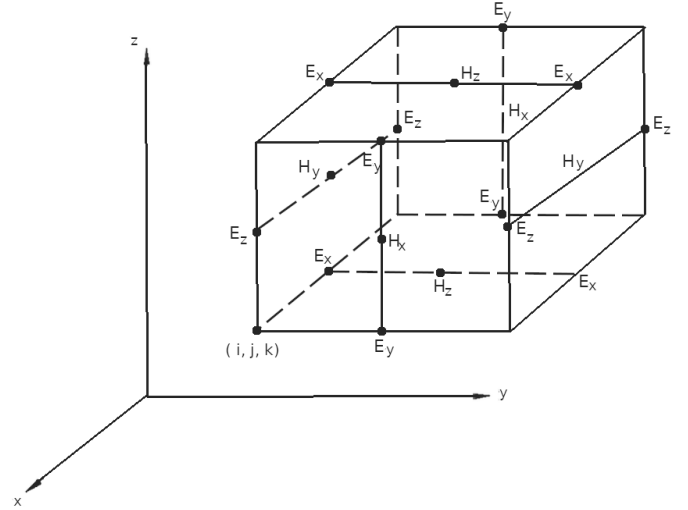}}
\caption{Yee lattice \cite{Yee}.}
\label{img:YeeLattice}
\end{figure}

Initially FDTD algorithm has been designed by Kane Yee \cite{Yee} as a technique to solve Maxwell equations in isotropic media. The fields are spatially distributed across a specific grid, so called
Yee lattice, shown in Fig. \ref{img:YeeLattice}. Time distribution is also different for electric and magnetic fields. The electric field stored at $k\cdot\Delta t$ time moments where $k$ is integer, while 
the magnetic fields keep $(k+\frac{1}{2})\cdot\Delta t$ time positions.
This pattern of fields allocation might seems strange, but shows robust stability and reduces required memory \cite{T1}.
\par As it has been shown at \cite{Yee} in isotropic non-dispersive case fields can be calculated as:
\begin{flalign}
\frac{B^{n+1/2}_{z}(i,j+\frac{1}{2},k+\frac{1}{2})-B^{n-1/2}_{z}(i,j+\frac{1}{2},k+\frac{1}{2})}{\Delta t}&= &&\\\nonumber 
\frac{E^{n}_{y}(i,j+\frac{1}{2},k+1)-E^{n}_{y}(i,j+\frac{1}{2},k)}{\Delta z}&- &&\\\nonumber 
\frac{E^{n}_{z}(i,j+1,k+\frac{1}{2})-E^{n}_{z}(i,j,k+\frac{1}{2})}{\Delta y}
\end{flalign}
The same way we obtain equations for $\vec D$ and $\vec H$ accounting also current $\vec J$.
\par When electric and magnetic inductions are known, we have to calculate the intensity fields ($\vec H$ and $\vec E$) for the next time step.
The way it is implemented in MEEP \cite{MEEP} can be assumed as a two step procedure. First we subtract material 
polarizations from the induction fields ($\vec D$ and $\vec B$), after we divide this value by $\varepsilon$ or $\mu$.
The used approach allows to account material responses of different physical nature at the same time. Each non-trivial material response corresponds to a polarization.
\section{Recursive convolutions}
The simplest case of magnetized plasma is a situation when cyclotron resonance is caused by Z-directed static magnetic field. The corresponding permittivity tensor can be found in \cite{ginzburg1962propagation}:

\begin{equation}\label{eps_tensor1}
\varepsilon(\omega)=\left(
\begin{matrix}
    \varepsilon_{xx} & j\varepsilon_{xy} & 0 \\
    j\varepsilon_{yx} & \varepsilon_{yy} & 0 \\
    0 & 0 & \varepsilon_{zz}
\end{matrix}
\right)
\end{equation}
where
\begin{align} \label{esp_chi}
\varepsilon_{xx}=\varepsilon_{yy}=1+\chi_{xx}=1+\chi_{yy}\\
\varepsilon_{zz}=1+\chi_{zz}\\
\varepsilon_{xy}=-\varepsilon_{yx}=\chi_{xy}
\end{align}
As it has been shown in \cite{Hunsb} the corresponding electric susceptibilities can be considered as real parts of complex variables:
\begin{align}\label{chi_xx_td_hat}
\hat\chi_{xx}(\tau)=\frac{\omega_p^2}{\nu_c^2+\omega_b^2}(\nu_c+j\omega_b)\left[1-e^{-(\nu_c-j\omega_b)\tau}\right]U(\tau)\\
\label{chi_xy_td_hat}
\hat\chi_{xy}(\tau)=\frac{\omega_p^2}{\nu_c^2+\omega_b^2}(\omega_b-j\nu_c)\left[1-e^{-(\nu_c-j\omega_b)\tau}\right]U(\tau)\\
\label{chi_zz_td_hat}
\hat\chi_{zz}(\tau)=\frac{\omega_p^2}{\nu_c}\left[1-e^{-\nu_c\tau} \right]U(\tau)
\end{align}
Corresponding polarization equations have been derived in the previous work \cite{our_paper1}:
\begin{equation}\label{P_splitting}
P^{(n)}_{i}= \vec P^{(n)}_{i, diag} + \vec P^{(n)}_{i, offdiag}
\end{equation}
where index $i$ can be $x$ or $y$.To treat polarizations as recursive convolutions and avoid expensive calculations, diagonal and off-diagonal parts are splitted into two parts each: 
\begin{equation}\label{conv2}
\begin{split}
P_{i,offdiag}=P_{i,offdiag,1} + P_{i,offdiag,2}
\end{split}
\end{equation}
Both of the terms of \ref{conv2} can be calculated as a recursive convolution:
\begin{equation}
\label{P1_rec}
P_{i,offdiag,1}^{(n+1)}=\vec E_{j}^{(n+1)}\frac{\omega_p^2}{\nu_c^2+\omega_b^2}(\omega_b-j\nu_c)\Delta t +P_{i,offdiag,1}^{(n)}\\
\end{equation}
\begin{equation}\label{P2_rec}
\begin{split}
P_{i,offdiag,2}^{(n+1)}=\frac{\omega_p^2}{\nu_c^2+\omega_b^2}\frac{\omega_b-j\nu_c}{j\omega_b-\nu_c}\left[E_{j}^{(n+1)} \left(e^{-(\nu_c-j\omega_b)\Delta t}-1\right) \right] +\\
+ P_{i,offdiag,2}^{(n)}\cdot e^{-(\nu_c-j\omega_b)\Delta t}
\end{split}
\end{equation}
Diagonal parts of the polarization calculated similar way and their recursive convolution form can be written as:
\begin{equation}\label{P_2xx}
\begin{split}
P_{i,diag,2}^{(n+1)}=E_{i}^{(n+1)}\frac{\omega_p^2}{\nu_c^2+\omega_b^2}\frac{\nu_c+j\omega_b}{j\omega_b-\nu_c}\cdot\\
\cdot\left(1-e^{-(\nu_c-j\omega_b)\Delta t}\right)+P_{i,diag,2}\cdot e^{-(\nu_c-j\omega_b)\Delta t}
\end{split}
\end{equation}
Material response to $E_z$ field is z-component of the polarization vector can be accounted a recursive convolution too:
\begin{equation}\label{P_zz_full}
P_{z}=P_{z,1}+ P_{z,2}
\end{equation}

\begin{equation}\label{P_1zz}
P_{z,1}^{(n+1)}=E_{z}^{(n+1)} \frac{\omega_p^2\Delta t}{\nu_c} + P_{z,1}^{(n)}
\end{equation}
\begin{equation}\label{P_2zz}
P_{z,2}^{(n+1)}=E_{z}^{(n+1)}\frac{\omega_p^2}{\nu_c^2}(e^{-\nu_c\Delta t}-1)+P_{z,2}^{(n)}\cdot e^{-\nu_c\Delta t}
\end{equation}
To distinguish physically valuable polarization from the parts, we will rename the parts and will call them "convolutions":
\begin{equation}
P_{diag/offdiag/z,1/2}=Conv_{diag/offdiag/z,1/2}
\end{equation}
\section{Coordinate frame change}
Consider two coordinate frames. The first $\vec r$ is our old coordinates where bias static magnetic field $\vec H^0$ directed along Z-axis. For simplicity sake we will consider $|\vec H^0|=1$.
 The second one $\vec r'$ is a new frame where the bias field is
arbitrary directed. There is a transformation matrix $T$ changing coordinates of vectors from $\vec r$ to $\vec r'$. As tensor $\varepsilon$ projects vector $\vec E(\vec r)$ into $\vec D(\vec r)$, in 
the new coordinate frame $\vec r'$ tensor $\varepsilon'$ should project $\vec E(\vec r')$ into $\vec D(\vec r')$:
\begin{equation}
\vec D(\vec r')=\varepsilon'\vec E(\vec r')
\end{equation}
If we rewrite $\vec E(\vec r')$ and $\vec D(\vec r')$ using $T$ matrix:
\begin{equation}
T\vec D(\vec r)=\varepsilon'T\vec E(\vec r)
\end{equation}
or 
\begin{equation}\label{proj1}
\varepsilon'=T\varepsilon T^{-1}
\end{equation}
Now we have to find $T$ matrix, to do this we can use Rodrigues' rotation formula \cite{Rodr}:
\begin{equation}\label{rodrf}
T=I+[v]_{\times} + [v]^2_{\times}\frac{1-c}{s^2}
\end{equation}
where $c$ is a cosine between the Z axis and the bias $\vec H$ field, s -- sinus os the same angle, $v=\vec H^{0} \times \vec e_z$ and $[v]_\times$ is defined as:
\begin{equation}
[v]_{\times}=\left(
\begin{matrix}
    0 & -v_3 & v_2 \\
    v_3 & 0 & -v_1 \\
    -v_2 & v_1 & 0
\end{matrix}
\right)
\end{equation}
or in our case this matrix can be written as: 

\begin{equation}
[v]_{\times}=\left(
\begin{matrix}
    0 & H^{0}_z & -H^{0}_y \\
    -H^{0}_z & 0 & H^{0}_x \\
    H^{0}_y & -H^{0}_x & 0
\end{matrix}
\right)
\end{equation}
while 
\begin{equation}\begin{split}
c=H^{0}_z\\
s=\sqrt{(H^{0}_x)^{2}+(H^{0}_y)^{2}}
\end{split}\end{equation}
The function (\ref{rodrf}) is not defined in a case when the sinus between $\vec H^0$ and the Z-axis is 0. To overcome this we manually define two cases. 
First, when $\vec H^0$ directed along the Z-axis, is trivial and the $T$ matrix is identity matrix. In the second case, when $\vec H^0$ is directed opposite the Z axis, described one of two possible matrices:
\begin{equation}\label{T_opp}
T=\left(
\begin{matrix}
    1 & 0 & 0 \\
    0 & -1 & 0 \\
    0 & 0 & -1
\end{matrix}
\right)
\end{equation}
\begin{equation}\label{T_opp2}
T=\left(
\begin{matrix}
    -1 & 0 & 0 \\
    0 & 1 & 0 \\
    0 & 0 & -1
\end{matrix}
\right)
\end{equation}
these matrices give the same result. For sake of certainty we will use (\ref{T_opp}).
\section{Implementation}
As elements of tensor $\chi$ are formulas one can not directly use (\ref{proj1}). Instead of that we can decompose $\chi$ as:
\begin{flalign}\label{chi_tensor1}
\chi&=\left(
\begin{matrix}
    \chi_{xx} & \chi_{xy} & 0 \\
    -\chi_{xy} & \chi_{yy} & 0 \\
    0 & 0 & \chi_{zz}
\end{matrix}
\right)=\chi_{xx}
\left(
\begin{matrix}
    1 & 0 & 0 \\
    0 & 1 & 0 \\
    0 & 0 & 0
\end{matrix}
\right)+ &&\\\nonumber 
&+\chi_{xy}
\left(
\begin{matrix}
    0 & 1 & 0 \\
    -1 & 0 & 0 \\
    0 & 0 & 0
\end{matrix}
\right)+\chi_{zz}
\left(
\begin{matrix}
    0 & 0 & 0 \\
    0 & 0 & 0 \\
    0 & 0 & 1
\end{matrix}
\right)=&&\\\nonumber 
&=\chi_{xx}M_{xx}+\chi_{xy}M_{xy}+\chi_{zz}M_{zz}
\end{flalign}
Applying (\ref{proj1}) on $M$ matrices from decomposition (\ref{chi_tensor1}) we obtain projection matrices:
\begin{equation}\label{Prmat}
\begin{split}
Pr_{diag}=T\cdot M_{xx}\cdot T^{-1}\\
Pr_{offdiag}=T\cdot M_{xy}\cdot T^{-1}\\
Pr_{z}=T\cdot M_{zz}\cdot T^{-1}
\end{split}
\end{equation}
The projection matrices (\ref{Prmat}) are generated in the constructor of "gyroelectric-susceptibility" class.
\par Since a high precision 3D simulation can require billions of grid cells, memory allocation is an important task. Memory allocation algorithm is determined in "new\_internal\_data" function of the class. 
The function allocates 2 complex double float numbers for each grid cell for every non-zero projection matrix element. Those recursive convolutions are calculated in "update\_P" function 
of "gyroelectric-susceptibility" class. New values of the polarizations are calculated as sum of products of convolutions and according projections:
\begin{flalign}
P_{i}(\vec r)&=\sum_j ( Pr_{ij,diag}\cdot Conv_{ij,diag}&&\\\nonumber 
&+Pr_{ij,offdiag}\cdot Conv_{ij,offdiag}+Pr_{ij,z}\cdot Conv_{ij,z} )
\end{flalign}
\par Function "subtract\_P" and the next steps are not affected by the change of coordinate system.
\section{Test case}
The analytical model is 9mm layer of magnetized plasma with the bias field $\vec H^0$ directed along the Z-axis.
Parameters for the analytical model are chosen to be: $\omega_p=2\pi\cdot 50\cdot 10^9$ rad/s, $\omega_b=2\pi\cdot 3\cdot10^{11}$ rad/s, and $\nu_c=2\cdot 10^{10}$ Hz.
As it has been shown in \cite{analytical_s} transmission of a slab can be written as:
\begin{equation}\label{T_1}
t=\frac{\tau_{12}\tau_{21} e^{-jh\cdot \hat n_2\cdot \omega /c}}{1+\rho_{21}\cdot\rho_{12} e^{-2jh\cdot \hat n_2\cdot \omega /c}}
\end{equation}
where $\rho$ and $\tau$ are the complex
Fresnel reflection and transmission amplitude coefficients for vacuum (index 1) and the plasma (index 2). 
\par
\begin{figure}[h]
\center{\includegraphics[width=0.8\linewidth]{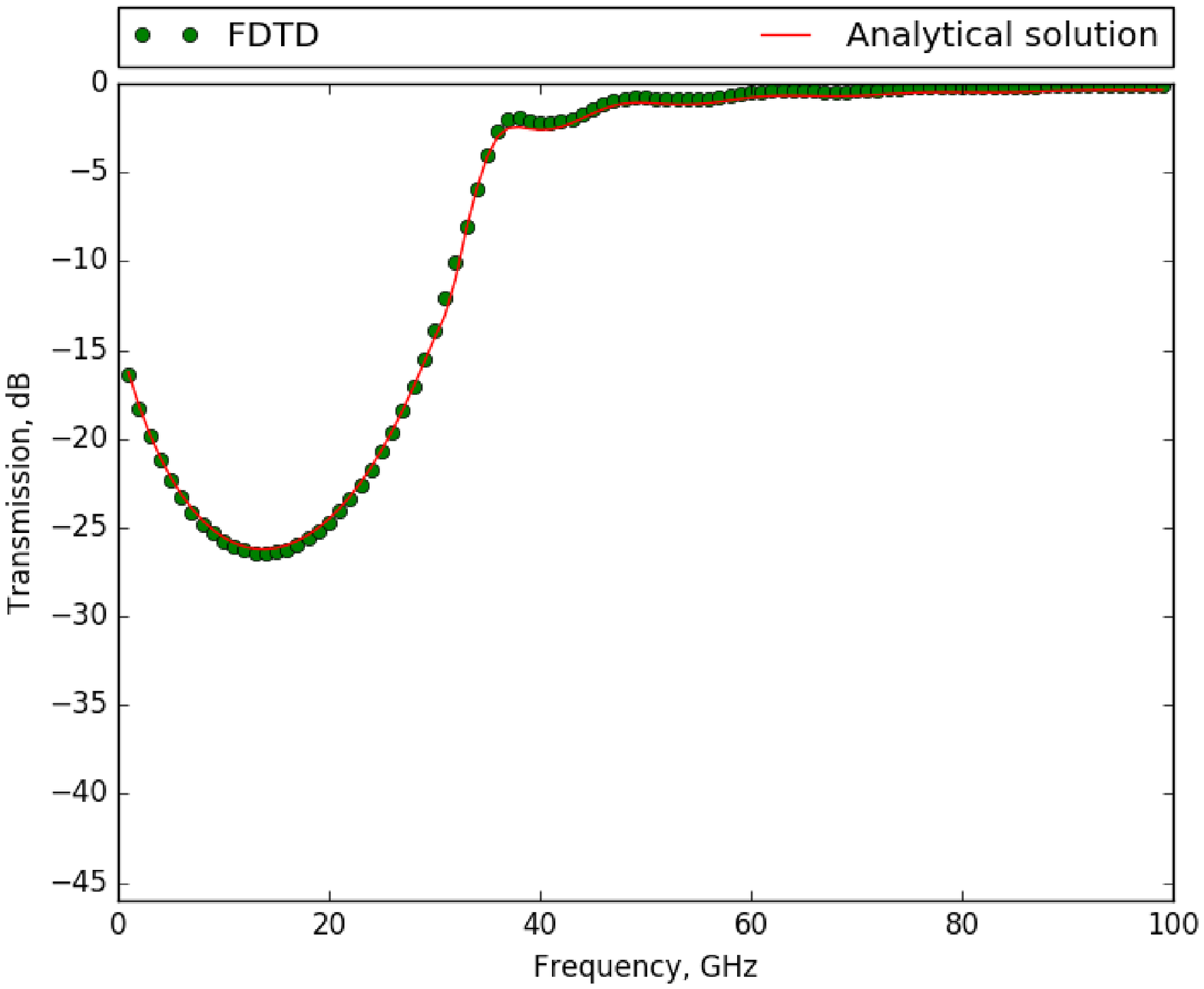}}
\caption{Comparison of the right circular polarized electromagnetic wave transmission.}
\label{img:RCP_t}
\end{figure}

\begin{figure}[h]
\center{\includegraphics[width=0.8\linewidth]{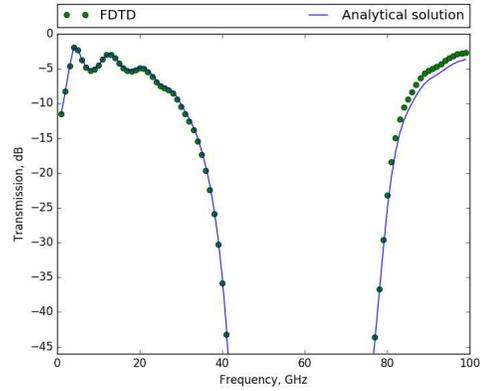}}
\caption{Comparison of the left circular polarized electromagnetic wave transmission.}
\label{img:LCP_t}
\end{figure}

 There are very strong dependence frequency of transmission $T=|t|^2$ for left and right circular polarized waves when we put according $n_2$ \cite{ginzburg1962propagation} into (\ref{T_1}):
\begin{equation}\label{n_rcp}
\hat n_{2RCP}^2=1- \frac{(\omega_p/\omega)^2}{1-j\nu_c/\omega - \omega_b/\omega}
\end{equation}
\begin{equation}\label{n_lcp}
\hat n_{2LCP}^2=1- \frac{(\omega_p/\omega)^2}{1-j\nu_c/\omega + \omega_b/\omega}
\end{equation}
The analytical solution is compared to the FDTD calculations result in Fig.\ref{img:RCP_t} and Fig.\ref{img:LCP_t}
\par To test the coordinate frame change algorithm the bias field $\vec H^0$ was directed opposite the Z-axis. In this case LCP and RCP transmission swapped as it should be according to the T matrix defined in 
(\ref{T_opp}). The reader might substitute it into equations (\ref{Prmat}) keeping in mind that in this case T matrix is involutory.
\section{Summary}
A new approach of FDTD simulation of magnetized plasma media allowing to set arbitrary direction of the bias $\vec H^0$ field is proposed. The implementation is based on MEEP, a widely used FDTD solver.
The simulation results for four cases compared to the analytical solutions and matched.
\par The source code is available on https://github.com/alexfriman/offidiagonal-meep
\section*{Acknowledgment}
The author would like to thank Nikolay Shubin for the discussion on linear algebra.

\end{document}